\documentclass[letterpaper]{article}
\usepackage{aaai}
\usepackage{times}
\usepackage{helvet}
\usepackage{courier}
\usepackage{graphicx}
\newcommand{\ignore}[1]{}
\begin{document}

\title{A Study of ``Churn'' in Tweets and Real-Time Search Queries\\(Extended Version)}

\author{Jimmy Lin \and Gilad Mishne\\
Twitter, Inc.\\
@lintool @gilad
}

\nocopyright

\maketitle
\begin{abstract}
\begin{quote}
The real-time nature of Twitter means that term distributions in
tweets and in search queries change rapidly:\ the most frequent
terms in one hour may look very different from those in
the next. Informally, we call this phenomenon ``churn''. Our
interest in analyzing churn stems from the perspective of real-time search. Nearly all
ranking functions, machine-learned or otherwise, depend on term statistics such as term frequency,
document frequency, as well as query frequencies. In the real-time context, how do we compute
these statistics, considering that the underlying distributions change
rapidly? In this paper, we present an analysis of tweet and query
churn on Twitter, as a first step to answering this question. Analyses
reveal interesting insights on the temporal dynamics of term
distributions on Twitter and hold implications for the design of
search systems.\\

This is an extended version of a similarly-titled paper at the
6th International AAAI Conference on Weblogs and Social Media (ICWSM 2012).
\end{quote}
\end{abstract}

\section{Introduction}

Twitter is a communications platform through which millions of users
around the world send short, 140-character tweets to their
followers. Particularly salient is the real-time nature of these
global conversations, which rapidly evolve to reflect breaking
events such as major earthquakes (e.g., Japan, March 2011), 
deaths of prominent figures (e.g., Steve Jobs, October 2011), or just
memes that idiosyncratically propagate across the internet. This paper
presents an analysis of the temporal dynamics of tweets and real-time
search queries. We focus specifically on the notion of ``churn'',
informally characterized as the process by which terms and queries
become prevalent and then ``drop out of the limelight''. To our
knowledge, this is the first large-scale study of this phenomenon.

We are interested in churn primarily from the perspective of real-time
search. A defining property of search in this context is the speed at which
relevance signals change, and thus a proper understanding and
treatment of this phenomenon is instrumental to search and derived
tasks (event tracking, query spelling correction, and so on).

Collection statistics such as {\it tf} (term frequency) and {\it df}
(document frequency) lie at the heart of any retrieval model more
complex then simple boolean retrieval. BM25~\cite{Robertson94},
language modeling~\cite{Ponte98}, as well as a myriad of other approaches
can be viewed as functional compositions of these basic statistics.
A modern learning-to-rank approach~\cite{LiHang_2011} typically uses dozens to
hundreds or more of query-document features derived from basic term statistics
(e.g., BM25 of specific index fields, incorporating term proximity features, etc.),
along with features capturing query statistics (e.g., frequency of query
{\it n}-grams within a query log), as well as query independent (i.e., document) features.
However, in real-time search, various collection and query statistics can change rapidly---frequencies
might increase or decrease orders of magnitude within a very short
time. What is the proper ``context'' to consider when trying to
compute term statistics for, say, BM25 or the language modeling
Dirichlet score? Computing term statistics across the entire
collection, as most retrieval models implicitly assume, does not seem
to be the proper treatment. The same problem manifests with query features:\
what should be the basis for computing query frequencies from a search log?
Finally, terms (both in tweets and in queries) are constantly being
introduced on Twitter (e.g., \#hashtags), some of which rise to
prominence within a short amount of time---this creates an
out-of-vocabulary problem when trying to compute various term
statistics.

This paper falls short of delivering solutions to the above
problems, but presents a characterization of the
phenomenon. We present analyses of churn on both tweets and real-time
search queries submitted to Twitter, with the hope that these results
will serve as the basis for tackling the search problem and that these
observations will be useful to the research community for other purposes as well.

\section{Methods}

\subsection{Metrics}

Let us begin with more precise definitions of the metrics we use to
quantify churn. We define a very simple measure of churn at rank $r$
as the fraction of terms in the top $r$ terms (as ordered by
frequency) at time interval $t_i$ that are no longer in the top $r$ at time
$t_j$. From this definition,
churn at rank $r$ of $0$ indicates the top $r$ terms are exactly the
same (but may be in different relative order), whereas at the opposite
end of the spectrum churn at rank $r$ of $1$ indicates that none of
the top $r$ terms are in common between the two time periods.

In our analyses, we consider different intervals (daily and hourly)
and different references points:\ in one set of experiments, we
consider interval-over-interval changes ($t_1$ vs. $t_2$, $t_2$
vs. $t_3$, $t_3$ vs. $t_4$, etc.); in another set of experiments, we
consider changes with respect to a fixed reference interval ($t_1$
vs. $t_2$, $t_1$ vs. $t_3$, $t_1$ vs. $t_4$, etc.).

This definition of churn does not capture changes in rank
within $r$, or the actual term frequencies.
To better characterize differences between term distributions
represented by successive time intervals, we use Kullback-Leibler (KL)
divergence, defined as follows:

\begin{displaymath}
D_{KL}(P || Q) = \sum_i P(i) \frac{P(i)}{Q(i)}
\end{displaymath}

\noindent Since KL divergence is not a symmetric measure, to be
precise, for the interval-over-interval condition,  we compute $D_{KL}(S_{t+1} || S_{t})$, where $S_{t}$ and
$S_{t+1}$ are the term distributions at time interval $t$ and $t+1$,
respectively. For the fixed reference condition, we compute $D_{KL}(S_{t} || S_{r})$, where
$S_{r}$ corresponds to the reference interval.
Following most information retrieval applications, we use base
2 for the log, resulting in an interpretation of the value as the
expected number of extra bits required to code samples from $P$ when
using a code based on $Q$, rather than using a code based on $P$.

To prevent the problem of zero probabilities (e.g., from
out-of-vocabulary terms), we smooth the maximum likelihood estimate
term probabilities using Bayesian smoothing with Dirichlet priors, as
follows:

\begin{displaymath}
P(w) = \frac{\textrm{c}(w) + \mu \cdot P_{bg}(w)}{\sum_t \textrm{c}(w) + \mu}
\end{displaymath}

\noindent where $\textrm{c}(w)$ is the count of term $w$ in the
appropriate window, $P_{bg}(w)$ is the background model, defined in
our case as the average of the distributions from the two intervals in question. Finally, $\mu$
is a smoothing hyperparameter (arbitrarily set to 10,000). A consequence of this
arbitrary setting is that absolute KL divergence values
are less meaningful than {\it comparisons} across different settings,
since absolute values are dependent on amount of smoothing
applied.

Finally, to quantify the impact of previously unseen terms, we compute
an out-of-vocabulary (OOV) rate, also defined in terms of a rank
$r$. An OOV rate at $r$ is the fraction of terms in the top $r$
(sorted by frequency) from time interval $t_i$ that is not observed
in the top $r$ during time interval $t_j$. An OOV rate of 0 means that
all top $r$ terms have been previously observed (hence, collection
statistics exist for them). A non-zero OOV rate means that
a retrieval engine must explicitly handle query terms that may
not exist in the collection (smoothing, backoff, defaults, etc.);
otherwise, the retrieval model may produce non-sensical results.

\subsection{Data and Processing}

Our analyses span the entire month of October 2011, both at the
daily and hourly level (all times are provided in UTC). We consider all
tweets created during that time, as well as all search queries
submitted to the twitter.com site. Since October has 31 days, this
corresponds to 30 different data points for the day-over-day analysis
and 743 data points for the hour-over-hour analysis. In the fixed reference
condition, we computed all statistics on a daily basis with respect to the first day
(30 data points) and on an hourly basis with respect to the first hour in the first day
(743 data points). Unfortunately, we
are unable to provide exact statistics on the size of our complete
dataset, except to note that according to publicly available figures (as of Fall 2011),
Twitter users create approximately a quarter of a billion tweets a
day, and Twitter search serves over two billion queries a day
(although this figure includes API requests).

For processing, Twitter's Hadoop-based analytics platform was used to
compute each distribution using Pig, a parallel dataflow language that
simplifies processing of large
datasets~\cite{Olston_etal_SIGMOD2008}. For the interested reader, a
recent paper provides details about Twitter's analytics
stack~\cite{Lin_etal_MAPREDUCE2011}. The resulting distributions
generated by Pig were fed into a custom program to compute the
various metrics described above.

\subsection*{Twitter's Trending Topics}
To highlight newsworthy topics gathering attention in tweets, Twitter
introduced Trending Topics:\footnote{support.twitter.com/entries/101125-about-trending-topics}\ a short list of
algorithmically-identified emerging trends and topics of discussion (both worldwide, and limited to certain geographic regions).
At its core, Twitter's trending topics algorithm is based on term
statistics:\ evidence for a term's recent prominence (e.g., appearances
in recent tweets, URLs, and so on) is continuously collected and
compared with longer-term statistics about the term. Terms (including
hashtags and phrases) that are significantly more prevalent in recent
data, compared with past data, are marked as trending topics and surfaced. Note that,
critically, phrases or hashtags become trending topics primarily
as a result of velocity (i.e., the rate of change in prevalence), {\it not}
sheer volume.

Trending topics are featured prominently both on the \mbox{twitter.com} site and on
Twitter clients, and a click on a trend leads to a Twitter search for
tweets containing the trend. This results in significantly elevated
query volumes for these terms and phrases during the time they were trending.
Often, during this period, these terms become top-searched queries.
Once they fall out of the list of top trends, due to a shift in
interest to other topics, their search volume declines rapidly and
they drop out of the list of top searches. Since this affects our
query churn observations, we repeat our analyses twice:\ once using all
queries, and once using all queries except those issued by clicking on
a trend.

\subsection{Churn in Web Queries}
To provide some context and to highlight differences in churn between
a real-time search engine and a general-purpose one, our results
should be compared to similar numbers obtained from web search
logs (from the same time period). Unfortunately, large-scale collections of web search queries are
generally not available for research purposes, with the exception of
the AOL query set~\cite{Pass:2006:PS:1146847.1146848}. As a point of
comparison, this corpus is far from ideal:\ it is relatively old
(2006); a sample of the full search stream; and drawn from a search
engine that no longer had dominant market position at the time the data was
collected.
For completeness, we report churn and OOV figures on this dataset,
but alert the reader to these caveats. Note that OOV
rates on sampled data are especially unreliable.
We omit the hourly
analyses from this source as the query frequencies are too low for
meaningful analysis (the top hourly terms are observed only a handful of
times).

\section{Results}

\begin{figure*}[p]
\centering\includegraphics[width=0.9\linewidth]{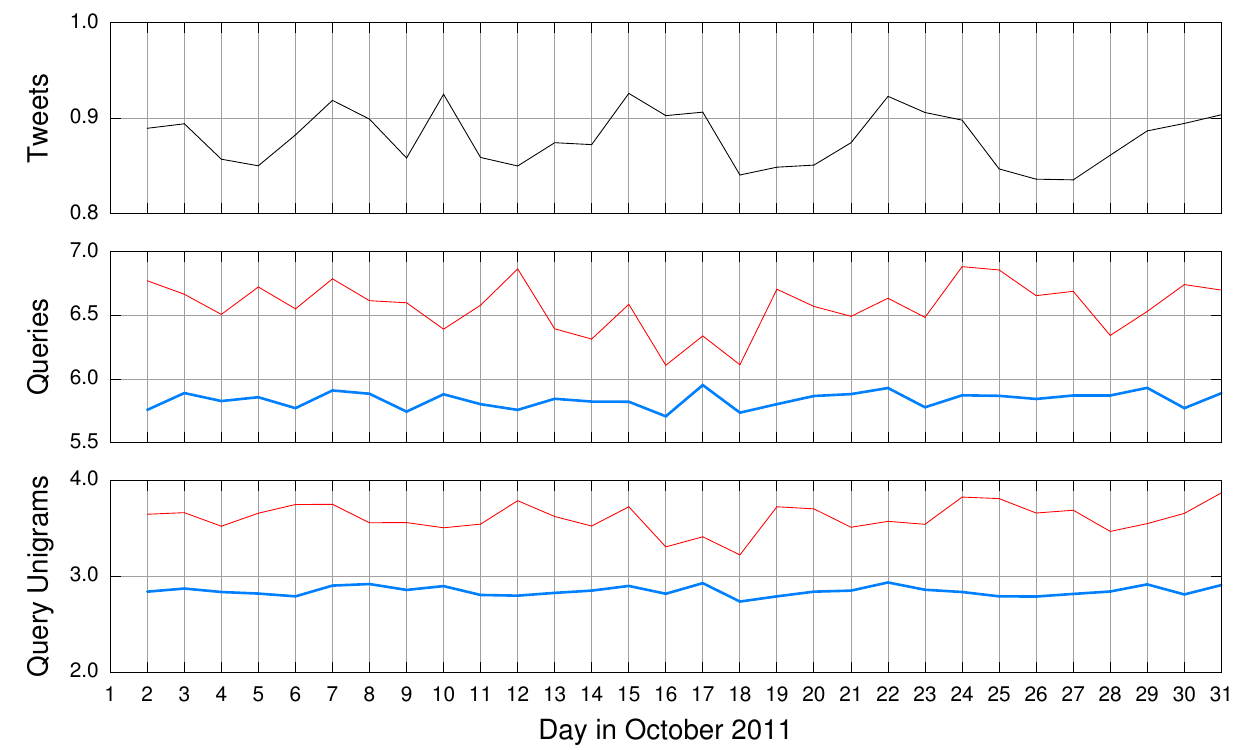}
\caption{Day-over-day KL divergence of tweets (top), queries (middle),
  and query unigrams (bottom). Middle and bottom graphs include
  analysis without trends (thick blue line) and with trends (thin
  red line).}
\label{figure:churn-kld-daily}
\end{figure*}

\begin{figure*}[p]
\centering\includegraphics[width=0.9\linewidth]{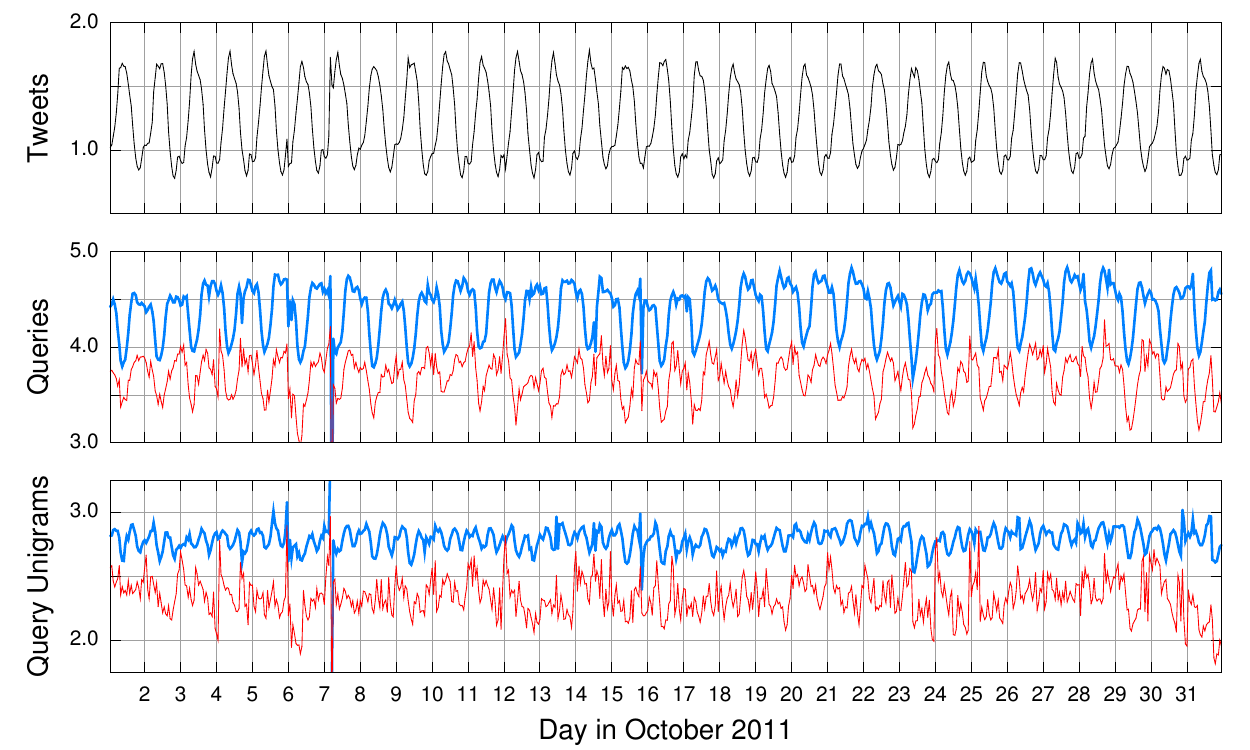}
\caption{Hour-over-hour KL divergence of tweets (top), queries (middle),
  and query unigrams (bottom). Middle and bottom graphs include
  analysis without trends (thick blue line) and with trends (thin
  red line).}
\label{figure:churn-kld-hourly}
\end{figure*}

\begin{table*}[t]
\begin{center}
\begin{tabular}{|l|rrrr||rrrr|}
\hline
 & \multicolumn{4}{|c||}{Churn Rate} & \multicolumn{4}{c|}{OOV Rate} \\
\cline{2-9}
                    & 10     & 100    & 1000   & 10000       & 10     & 100    & 1000   & 10000 \\
\hline
\hline
Tweets              & 0.0167 & 0.0233 & 0.0407 & 0.0682      & 0.0000 & 0.0000 & 0.0008 & 0.0012 \\
Queries             & 0.8067 & 0.8180 & 0.6413 & 0.3199      & 0.4500 & 0.4073 & 0.2678 & 0.0702 \\
Queries ($-$T)      & 0.4433 & 0.3807 & 0.3166 & 0.2722      & 0.0400 & 0.0450 & 0.0317 & 0.0313 \\
Q.\ Unigrams        & 0.8067 & 0.6937 & 0.4105 & 0.1754      & 0.1633 & 0.0960 & 0.0732 & 0.0297 \\
Q.\ Unigrams ($-$T) & 0.2500 & 0.1360 & 0.1254 & 0.1319      & 0.0100 & 0.0043 & 0.0051 & 0.0070 \\
\hline
Web Queries         & 0.1107 & 0.2139 & 0.3670 & 0.7584      & 0.0000 & 0.0000 & 0.0290 & 0.5402 \\
Web Q. Unigrams     & 0.0410 & 0.1608 & 0.1740 & 0.3401      & 0.0000 & 0.0000 & 0.0570 & 0.1641 \\
\hline
\end{tabular}
\end{center}
\caption{Results of day-over-day analysis, showing query churn (left
  half of the table) and OOV rate (right half of the table) at various
  ranks $r$ for:\ tweets, all queries, all queries minus trends, all
  query unigrams, all query unigrams minus trends, and web queries
  from the AOL dataset.}
\label{table:churn-daily}
\end{table*}

\begin{table*}[t]
\begin{center}
\begin{tabular}{|l|rrrr||rrrr|}
\hline
 & \multicolumn{4}{|c||}{Churn Rate} & \multicolumn{4}{c|}{OOV Rate} \\
\cline{2-9}
 & 10 & 100 & 1000 & 10000 & 10 & 100 & 1000 & 10000 \\
\hline
\hline
Tweets              & 0.0349 & 0.0349 & 0.0653 & 0.1004      & 0.0000 & 0.0000 & 0.0000 & 0.0015 \\
Queries             & 0.3055 & 0.3094 & 0.2880 & 0.2880      & 0.0424 & 0.0967 & 0.0452 & 0.2479 \\
Queries ($-$T)      & 0.2899 & 0.2768 & 0.3336 & 0.5281      & 0.0096 & 0.0158 & 0.0329 & 0.2599 \\
Q.\ Unigrams        & 0.3257 & 0.2930 & 0.1687 & 0.3059      & 0.0073 & 0.0302 & 0.0137 & 0.0598 \\
Q.\ Unigrams ($-$T) & 0.1783 & 0.1639 & 0.1657 & 0.3144      & 0.0027 & 0.0019 & 0.0032 & 0.0622 \\
\hline
\end{tabular}
\end{center}
\caption{Results of hour-over-hour analysis, showing query churn (left
  half of the table) and OOV rate (right half of the table) at various
  ranks $r$ for:\ tweets, all queries, all queries minus trends, all
  query unigrams, all query unigrams minus trends.}
\label{table:churn-hourly}
\end{table*}

The day-over-day analysis in terms of KL divergence is shown in
Figure~\ref{figure:churn-kld-daily}. The figure is broken into three
graphs:\ results over tweets (top), queries (middle), and query
unigrams (bottom). The difference between the last two is worth
explaining:\ for analysis in terms of queries, we consider the entire
query string (which might consist of multiple terms) as a distinct
event. This would, for example, consider ``steve jobs'' and ``jobs''
distinct events. For analysis in terms of query unigrams, all queries
are tokenized into individual terms, and we consider the multinomial
distribution over the term space. Both analyses are useful:\ when
building a query model, for example~\cite{Zhai_Lafferty_SIGIR2002},
or extracting query-level features for learning to rank, estimates over the
event space of queries would yield more signal, although due to
sparsity, one would typically need to back off to unigram statistics. For both
the middle and bottom graphs in Figure~\ref{figure:churn-kld-daily},
we further break down analysis in terms of all queries (thin red line)
and with trends discarded (thick blue line).
We make a few observations:\ First, there does not appear to be cyclic
patterns in day-over-day churn (e.g., day of week effects). Second,
eliminating trend queries reduces KL divergence, i.e., successive
distributions appear more ``similar''---this makes complete sense given the
nature of trending topics.

The hour-over-hour analysis in term of KL divergence is shown in
Figure~\ref{figure:churn-kld-hourly}:\ this graph is organized in
exactly the same way as Figure~\ref{figure:churn-kld-daily}. For
tweets and queries, but not query unigrams, we observe strong daily
cyclic affects. This is driven by a combination of the rhythm of
users' daily activities and the international nature of Twitter:\ for
example, as users in the United States go to bed and users in Japan
wake up, the composition of tweets and queries naturally changes,
resulting in churn. Interestingly, we observe that removing trends
actually {\it increases} churn, i.e., we observe higher KL divergence
values. This suggests that the typical lifespan of a
trending topic is longer than an hour, i.e., trending topics churn
``naturally'' at a rate that is slower than hourly, such that removing
those events from the distribution increases overall churn. The time
that a particular topic trends is a function of many factors:\ for
news, factors include significance of the news event and interactions
with competing stories; for internet memes, the lifespan of a
particular hashtag is often idiosyncratic. Although not in the Twitter
context, see~\cite{Leskovec:2009:MDN:1557019.1557077} for a quantitative analysis of this ``news
cycle''.

Table~\ref{table:churn-daily} presents churn rates and OOV rates at
rank $r=\{10, 100, 1000, 10000\}$ for the day-over-day analysis,
averaged across the entire month, for all five experimental
conditions:\ tweets, queries, and query unigrams ($\pm$ trends for the
last two). Table~\ref{table:churn-hourly} shows similar results for
the hour-over-hour analysis. 

From the search perspective, the OOV
rates are interesting, in that they highlight a challenge that
real-time search engines must contend with. Take, for example, the
day-over-day query unigram OOV rate at rank 1000:\ results tell us that
7.32\% of query terms were not observed in the previous day. This means
that for a non-trivial fraction of query unigrams, we have no
query-level features:\ query frequency, clickthrough data to learn
from, etc. This is the result at rank 1000, which represents queries
pretty close to the head of the distribution. Of course, this
particular analysis includes trends (and removing trends reduces the
OOV rate substantially), but trend queries remain an important class
of queries for which we would like to return high quality results.

\begin{figure*}[p]
\centering\includegraphics[width=0.9\linewidth]{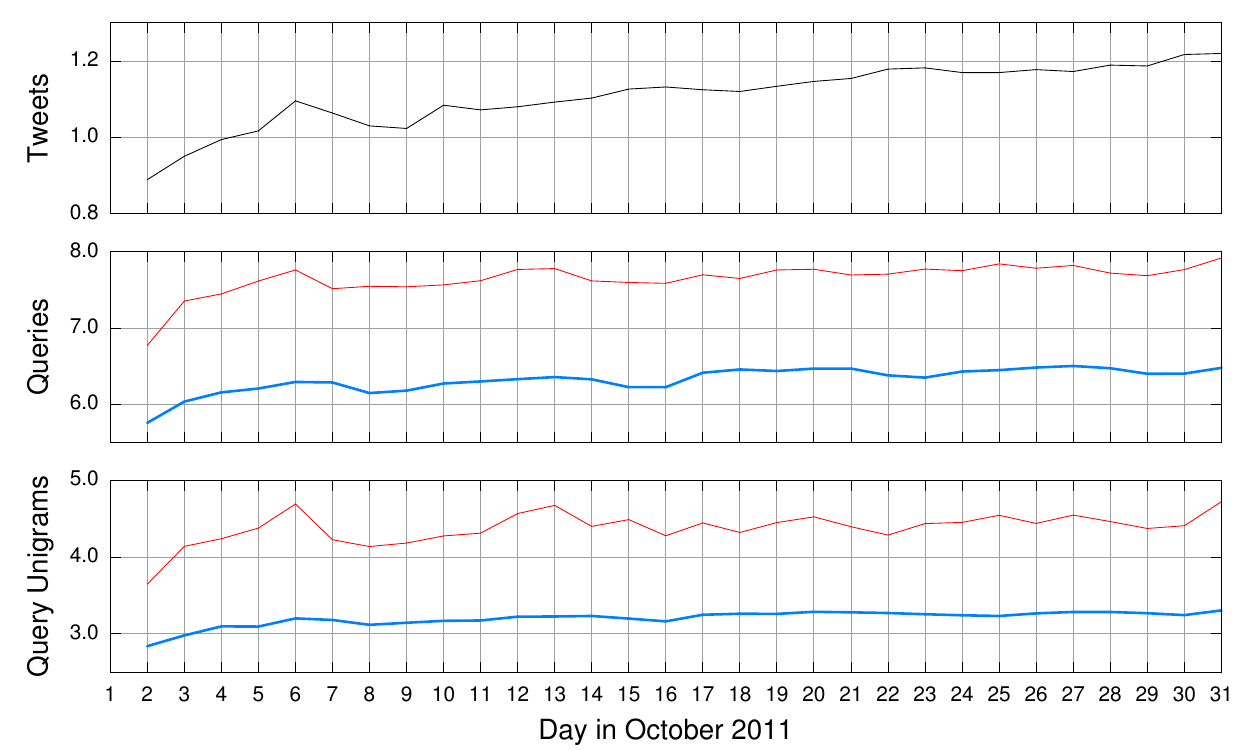}
\caption{KL divergence with respect to first day of October of tweets (top), queries (middle),
  and query unigrams (bottom). Middle and bottom graphs include
  analysis without trends (thick blue line) and with trends (thin red
  line).}
\label{figure:fixed-ref:churn-kld-daily}
\end{figure*}

\begin{figure*}[p]
\centering\includegraphics[width=0.9\linewidth]{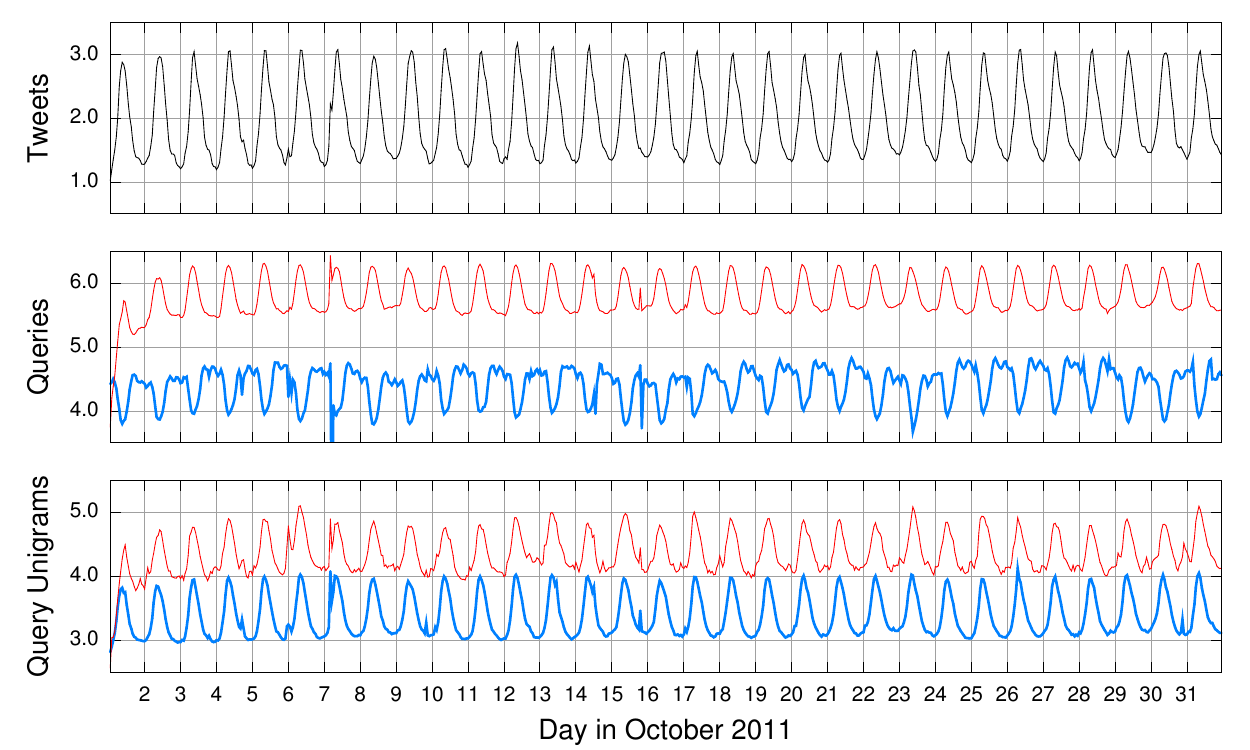}
\caption{KL divergence with respect to first hour in October of tweets (top), queries (middle),
  and query unigrams (bottom). Middle and bottom graphs include
  analysis without trends (thick blue line) and with trends (thin red
  line).}
\label{figure:fixed-ref:churn-kld-hourly}
\end{figure*}

In addition to the interval-over-interval analysis, we repeated the
same exact set of experiments with respect to a fixed
reference. Figure~\ref{figure:fixed-ref:churn-kld-daily} shows
KL divergence of every day in October with respect to the first
day. Figure~\ref{figure:fixed-ref:churn-kld-hourly} shows
KL divergence of every hour in October with respect to the first hour.
In the daily condition, we observe weak cyclic effects (i.e., day of
week cycles). In Figure~\ref{figure:fixed-ref:churn-kld-daily}, for
most of the plots we observe a ``trough'' approximately a week after
the initial day.
In the hourly condition, we observe strong cyclic affects, as we would expect.

\subsection{Zooming In}
In the context of term churn, rapidly-unfolding events such as natural
disasters or political unrest are of particular interest. In such
scenarios term frequencies may change significantly over short periods
of time as the discussion evolves. Our next analysis examines one 
such event, the death of Steve Jobs, the co-founder and CEO of Apple,
in the afternoon hours of October 5th (around midnight UTC).

\begin{figure*}[p]
\centering\includegraphics[trim=0cm 0cm 0cm 4.75cm, clip=true, width=0.9\linewidth]{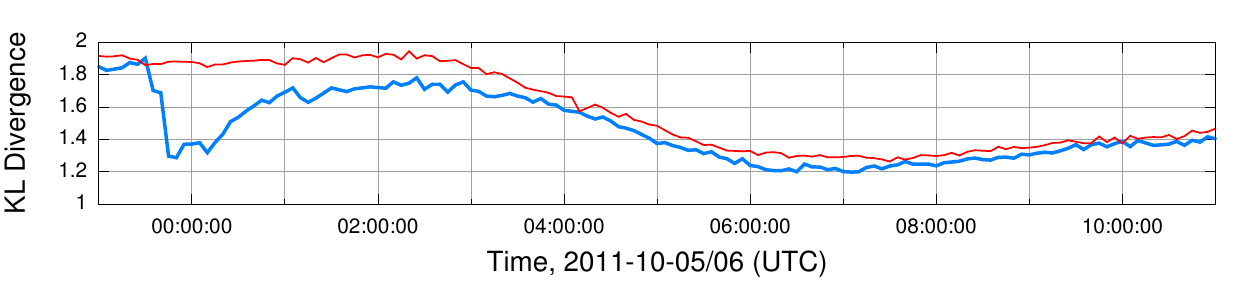}
\vspace{-0.2cm}
\caption{KL divergence of the query stream, in intervals of 5 minutes, over a
12 hour period during a major event (in blue, thick line); the overlay (red, thin)
shows the 5-minute KL divergence during the same hours in the preceding day, for reference.}
\label{figure:five-min-churn}
\end{figure*}

Figure~\ref{figure:five-min-churn} shows the KL divergence at 5-minute
intervals over a period of 12 hours surrounding the event (thick blue line), contrasting
it with the 5-minute KL divergence values over the same hours in the 
previous day (thin red line). Note the sharp drop in divergence as the real-time query
stream focuses on the event. A few hours later, divergence converges
to a pattern close to the previous day, although actual
values are around 10\% less, as significant portions of the query stream
continue to discuss the event.

\begin{figure*}[p]
\vspace{-3cm}
\centering\includegraphics[trim=0cm 0cm 0cm 2.4cm, clip=true, width=0.9\linewidth]{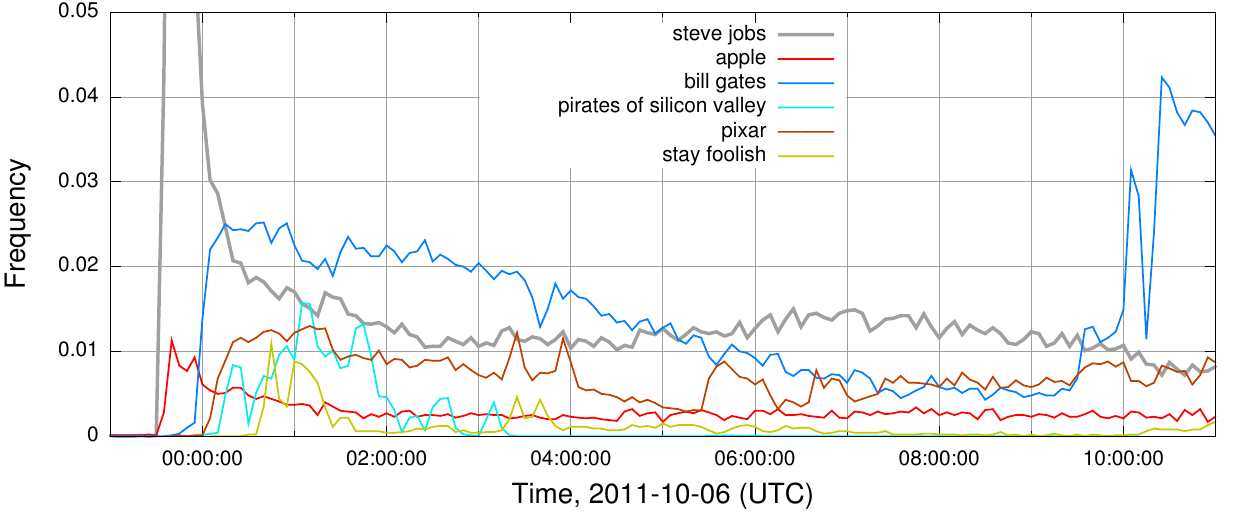}
\vspace{-0.2cm}
\caption{Frequencies of queries related to Steve Jobs' death over a 12 hour period
in 5-minute intervals, normalized to the total number of queries in the interval.
At its peak, the query ``steve jobs'' reaches 0.15 (15\% of the query stream); for
readability of the other query frequencies, the scale is not stretched to include this point.}
\label{figure:stevejobs-5-min}
\end{figure*}

However, even within a particular event, churn of individual queries and 
terms vary. Figure~\ref{figure:stevejobs-5-min} shows the frequency
of several queries related to the event over the same time period, in
5-minute intervals. Note that the patterns do not necessarily correlate in
timespan or shape, displaying a range of exponential, linear, and irregular
decays over time. One possible conclusion from this data is that a
simple approach to account for changes in term frequency over time, such as
applying a decay function over the frequency, may not be sufficient. Additionally,
it is clear that, at least during major events, sub-hour updates to various collection
and query term statistics are essential.

\begin{table*}[p]
\vspace{-2.75cm}
\begin{center}
\begin{tabular}{|c|lc||lc|}
\hline
 & \multicolumn{2}{|c||}{Tweets} & \multicolumn{2}{c|}{Queries} \\
\hline
Date & Term & Frequency Ratio & Term & Frequency Ratio \\
\hline
2011-10-03 & sunday & 3.61 & knox & 3.45  \\
2011-10-04 & weekend & 1.62 & iphone & 4.66 \\
2011-10-05 & iphone & 1.48 & iphone & 1.62  \\
2011-10-06 & steve & 8.28 & \#stevejobs & 7.84  \\
2011-10-09 & sunday & 3.46 & barlow & 2.82 \\
2011-10-13 & blackberry & 3.57 & guetta & 8.77 \\
\hline
\end{tabular}
\end{center}
\caption{A sample of top churning terms in tweets (left) and the query stream (right). For each term, the date it was a top term is shown (UTC), as well as the ratio between its frequency on a that day and on the following one.}
\label{table:top-churning}
\end{table*}

\subsection{Churning Terms}
Finally, we briefly examine the terms that cause high churn rates in tweets and
in the query stream. Table~\ref{table:top-churning} shows unigrams taken from
the set of terms that were among the top 500 terms in tweets and queries
on one day, and not in the top 2000 terms in the following day; for each day sampled,
we show the top non-stopword term.

One interesting observation from an anecdotal examination of the data is that term
churn in tweets appears more cyclic and predictable than that in the query stream.
While churn levels are high in both, in the case of the query stream this is largely
driven by news events, whereas in tweets this is driven by a combination of shifting
cyclical interests (e.g., weekends) and news. This suggests that
while both tweets and queries experience large amounts of churn, they are
{\it qualitatively} different. For example, terms like ``weekend'' and ``party'' in tweets have frequencies
that rise and fall predictably, whereas variations in query frequencies
appear to be far less predictable.

\section{Related Work}

In the domain of temporal information retrieval, a large body of work on timestamped
corpora was driven by TDT---the Topic Detection and Tracking
initiative~\cite{Allan:2002:TDT:772260}. Among the findings relevant
to our work is the demonstration that document retrieval benefits from
incorporating temporal aspects of the collection into the ranking,
e.g., via a language model~\cite{Li:2003:TLM:956863.956951} or by
utilizing additional statistics about term frequencies over time,
rather than just a global
weight~\cite{Efron:2010:LTS:1855138.1855147}. Retrieval can also
be improved by taking into account
the temporal distribution of results and modeling
the ``burstiness'' of events~\cite{JonesRosie_Diaz_TOIS2007}.
 
A body of recent work focuses on the temporal dynamics of Twitter {\em
  content} (rather than search queries).  For example, Petrovi\'{c} et
al.~\shortcite{Petrovic:2010:SFS:1857999.1858020} apply a TDT
task---first story detection---to Twitter data, while Wu et
al.~\shortcite{DBLP:conf/icwsm/WuTKM11} develop predictors for term
churn in tweets.
 
In the context of web search, Beitzel et
al.~\shortcite{Beitzel:2004:HAV:1008992.1009048} track the relative
popularity of queries throughout the day, as well as the hourly
overlap between queries. While exact figures vary by query category,
they observe relatively high correlations between the queries of a
given hour and the hour following it, with few exceptions related to
breaking news. A later study by Kulkarni et
al.~\shortcite{Kulkarni:2011:UTQ:1935826.1935862} groups queries by
the shape of their popularity over time, showing significant
differences over time. However, they focus on a small set of
manually-selected queries rather than examining properties of the full
query stream.
 
Of particular interest to us is the work of Teevan et
al.~\shortcite{Teevan:2011:TCM:1935826.1935842}, who analyze several
aspects of Twitter queries and compare them to the web query
stream. Interestingly, they observe a {\em lower} churn rate on
Twitter than on the web. This is counter-intuitive, and may be attributed
to the relatively limited amount of data analyzed (our collection is
several orders of magnitude larger, as well as annotated for the
presence of trends, cleaned of spam, and so on).
 
Also related to our study is the extensive
analysis of modifications to web documents over time, presented
in~\cite{Adar:2009:WCE:1498759.1498837}. The types of change patterns
we observe appear very different from the patterns the authors
identify on web pages (e.g., ``hockey stick'' curves), which
naturally makes sense given the context.

\section{Conclusions}

This paper examines changes to term distributions on Twitter over
time, focusing on the query stream and the implications for ranking in
a real-time search system. Our observations can be summarized as
follows:

\begin{itemize}
\item {\bf Churn.} Term distributions change rapidly---significantly faster than in web search for the head of the distribution. Even after discounting trending terms promoted by the platform, churn rates of top real-time queries are up to four times higher than those of web searches. For the tail of the distribution, churn drops quickly, and appears to be lower than that observed in web queries.
\item {\bf Unobserved terms.} Similarly, rates of out-of-vocabulary words are higher for top Twitter queries, but lower at the tail of the distribution. This translates to rapid changes in the top user interests, but relative stability in the topics for which users seek real-time results.
\item {\bf Update frequency.} Although query churn is consistently high, during major events it can further increase dramatically, as queries change minute by minute. In fact, to maintain accurate collection statistics requires frequent term count updates---in intervals of 5 minutes or less, according to our data.
\item {\bf Churn patterns.} The time period in which a query remains a top one varies, as does its decay pattern;
{na\"{i}ve} approaches such as fixed term frequency decays may not be able to correctly model frequency
changes over time.
\item {\bf Predictability.} Anecdotal evidence suggests that some query churn may be predicted from past observations, providing a potential source for addressing this issue.
\end{itemize}

The growing importance of real-time search brings with it several
information retrieval challenges; this paper frames one such challenge, that of
rapid changes to term distributions, particularly for
queries. In follow-up work we plan to evaluate techniques for handling
the volatility of the real-time search stream and the limited
collection statistics that exist for new queries.

\bibliographystyle{aaai} 
\bibliography{churn}

\end{document}